\documentclass[sigconf]{acmart}
\usepackage{graphicx}
\usepackage{comment}
\usepackage{threeparttable}

\AtBeginDocument{%
  \providecommand\BibTeX{{%
    \normalfont B\kern-0.5em{\scshape i\kern-0.25em b}\kern-0.8em\TeX}}}

\begin{document}

\title{Accelerated learning from recommender systems using multi-armed bandit}


\author{Meisam Hejazinia}
\email{mnia@expediagroup.com}
\author{Kyler Eastman}
\email{keastman@expediagroup.com}
\author{Shuqin Ye}
\email{sye@expediagroup.com}
\affiliation{%
  \institution{Vrbo, part of Expedia Group}
}
\author{Abbas Amirabadi}
\email{aamirabadi@expediagroup.com}
\author{Ravi Divvela}
\email{rdivvela@expediagroup.com}
\affiliation{%
  \institution{Vrbo, part of Expedia Group}
}

\begin{abstract}
Recommendation systems are a vital component of many online marketplaces, where there are often millions of items to potentially present to users who have a wide variety of wants or needs. Evaluating recommender system algorithms is a hard task, given all the inherent bias in the data, and successful companies must be able to rapidly iterate on their solution to maintain their competitive advantage. The gold standard for evaluating recommendation algorithms has been the A/B test since it is an unbiased way to estimate how well one or more algorithms compare in the real world. However, there are a number of issues with A/B testing that make it impractical to be the sole method of testing, including long lead time, and high cost of exploration. We argue that multi armed bandit (MAB) testing as a solution to these issues. We showcase how we implemented a MAB solution as an extra step between offline and online A/B testing in a production system. We present the result of our experiment and compare all the offline, MAB, and online A/B tests metrics for our use case. 

\end{abstract}

\keywords{multi armed bandit, recommender system, embedding, A/B testing, vacation rental}


\maketitle

\section{Introduction}

Recommendation systems are a vital component of many online marketplaces, where there are often millions of items to potentially present to users who have a wide variety of wants or needs. Determining each item's relevance to each user can be a hard algorithmic problem, and successful companies must be able to rapidly iterate on their solution to maintain their competitive advantage.

The gold standard for evaluating recommendation algorithms has been the A/B test since it is an unbiased way to estimate how well one or more algorithms compare in the real world. However, there are a number of issues with A/B testing that make it impractical to be the sole method of testing.  First of all, gathering enough traffic to reach statistical significance can take unreasonably long, especially when the full shopping cycle can last days or weeks. When there are multiple algorithms to test, one can either test each algorithm in series or as multiple variants in parallel.  However, neither solution significantly cuts down on the iteration time, with multiple variants taking longer due to the smaller amount of traffic within each variant. 

Another problem with online A/B testing is that it necessitates showing potentially sub-optimal algorithms to the real-world marketplace, which can ultimately cost companies money by delivering a degraded experience to users by presenting them with irrelevant products. This can provide a perverse incentive for algorithm developers to test incremental improvements.  Another option is to reduce the traffic to a variant that is considered "risky".  However, such a test will again take longer to reach statistical significance. 

Because of the required time and potential risks of online A/B testing, one can potentially make decisions about algorithms beforehand through offline evaluation metrics.  However, these metrics can be inherently biased by the current algorithm or the user interface. Since the algorithm determines when and how an item is presented, one cannot determine that item's relevancy to users apart from the algorithm's estimate. An item might have a higher relevance score if presented according to a different algorithm. In fact, these biases can be remedied if we leverage a randomize or stochastic algorithm.

Multi-armed Bandit tests bridge the gap between quick but biased offline metrics and unbiased but long online A/B testing.  Like an A/B test, a MAB test starts with equal traffic devoted to each variant. However, that traffic allocation is changed after a short period according to the particular MAB design, increasing or decreasing traffic according to each variant's performance and taking into account uncertainty for each variant in a principled way. This way, traffic is not wasted on precisely determining the level to which some variants under perform. Instead, more traffic is devoted to high performing variants increasing the ability to statistically determine the difference between them. Furthermore, this reallocation minimizes the exposure of users to sub-optimal algorithms. 

While MAB testing can save time over traditional A/B testing, we have found in practice it cannot completely replace it. MAB tests require a metric that is sensitive to change in a short period of time, like click-though rate, which may not correlate with a longer term metric used in the evaluation of an AB test. Instead, it can be an important step to validate offline metrics with some initial unbiased results, while allowing the opportunity to refine the candidate algorithms for a further A/B test. A typical development cycle is illustrated in figure 1.  Model development, offline simulation, and integrating into a production environment typically can take 4-6 weeks, and can be done asynchronously across multiple contributors.  A MAB test on multiple variants (typically up to 6) will then take about one to two weeks. The winner of the MAB test will then get tested in an A/B test against the current algorithm, which takes about 4 weeks.  With this hybrid MAB/AB approach, one could test up to six algorithms within a five week period. This is possible by eliminating candidates that grossly under-perform on short-term metrics.  Alternatively, a sole A/B testing protocol would take 6*4weeks = six months.  This allows for each A/B test to be much more productive, each test validating a more cultivated set of alternative algorithms. However, as we will discuss more in-depth, our decision to promote an algorithm to the next step is still done on a case-by-case basis, integrating information across all steps and a variety of metrics, including the results of previous tests. 

\begin{figure}[ht]
\centering
\includegraphics[width=\linewidth]{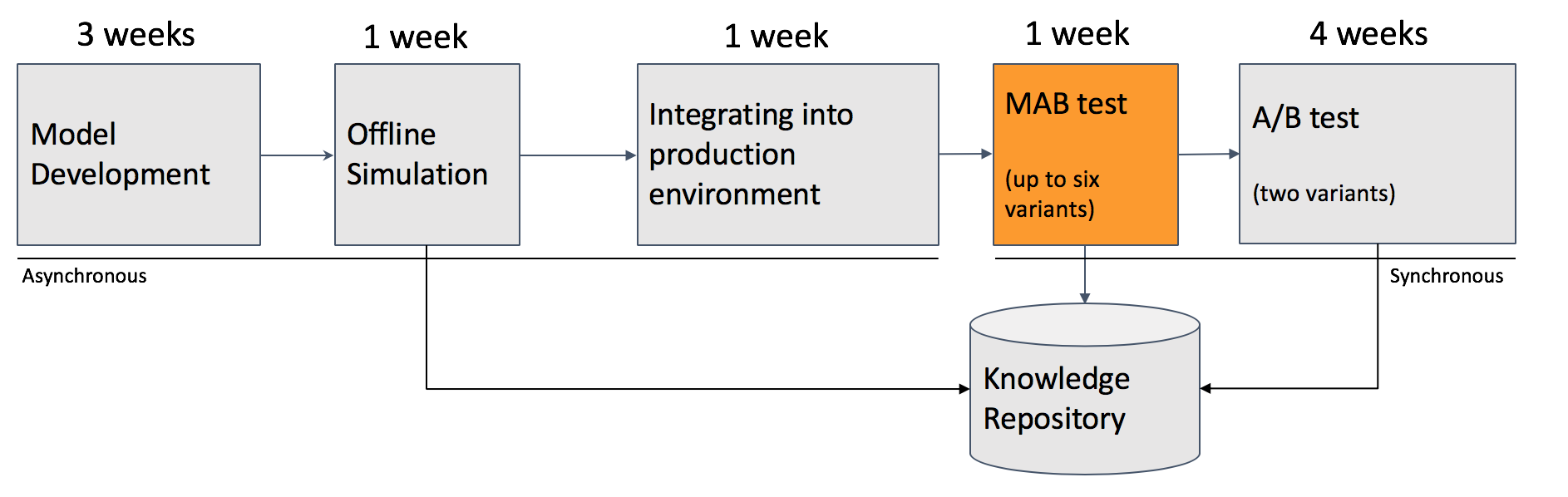}
\caption{Mult-armed Bandit testing in the larger context of evaluating recommendation systems}
\end{figure}

The rest of the paper is organized as follows. We first describe the current state of related work (section 2), discussing both theoretical studies of MAB testing frameworks and empirical studies of MAB tests. We then review the current state of recommendation algorithms, and the MAB architecture (section 4).  We then review the architecture developed for running our MAB test, and describe the models we test in this case study. We then present the results of our case study in Section 5 and 6, with offline metrics, a MAB test, and an online test. We then describe the implications of the findings in the discussion section (section 7). We then conclude in section 8.

\section{Related works}

Academics have extensively studied algorithms that solve the multi-armed bandit problem of balancing the acquisition of new knowledge (exploration) and leveraging the current knowledge (exploitation) when making decisions under uncertainty. At the same time, a number of companies have described how they leverage multi-armed bandits for making recommendations.  

\subsection{Theoretical Studies of Multi-armed Bandits}
Most theoretical studies propose algorithms that can be shown to have a minimal asymptotic regret bound \cite{kveton2019perturbed,cesa2017boltzmann,garivier2018kl,cappe2013kullback,rosenski2016multi}. Some studies suggest deterministic algorithms such as Upper Confidence Bound (UCB) and its variants \cite{garivier2011kl,liu2018change,kwon2017sparse,liu2018ucboost,lattimore2015optimally,lattimore2016regret,filippi2010optimism}, since they have a feasible closed-form index policy. Others promote Stochastic Bayesian Thompson sampling variants, since they have practical advantages, since it does not get stuck in the same action over extended period of time under delayed feedback \cite{raj2017taming,bubeck2012regret,combes2017minimal}. However, these studies often assume environments to be stationary and feedback to be independent and identically distributed (IID), assumptions that we already know to be false in our case.   

Some studies have addressed how to make an algorithm robust to non-stationary and non-IID environments. Some studies have focused on developing robust or distributed algorithms for the non-cooperating adversarial settings, where multiple players engage in the zero-sum games \cite{audibert2010regret,besson2017multi,evirgen2017effect,kim2019optimality,zimmert2018optimal,garivier2016explore,singla2017learning,avner2014concurrent,liu2010distributed,boursier2018sic,lugosi2018multiplayer,seldin2017improved,kalathil2014decentralized,ProductizationContextualBandit20191}.  Others solve the issue, not by conditioning on the context, but by defining sliding windows, piece-wise stationary assumptions, and discounting processes \cite{agarwal2016corralling,cao2018nearly,moulines19858,wei2018abruptly}. Other authors in academia have studied contextual multi-armed bandit and its variants regret bounds \cite{zeng2016online,lomas2016interface,ahonen2017applying,mcconachie2017bandit}.

\subsection{Empirical Evaluation of Recommendation Algorithms}
The variant of multi-armed bandit that is popular in industry is contextual bandit, for its ability to handle cold start problem at scale \cite{mcinerney2018explore}. Cold start problem refers to the state that recommendation system has not gathered enough data to draw inference about a user or an item. Many studies highlight how vanilla bandit can be extended to contextual multi-armed bandit solutions for personalization and recommendation systems \cite{chapelle2011empirical,mao2018batched,li2010contextual,li2012unbiased,agarwal2014taming,ross2013normalized}. For example, a study \cite{ProductizationContextualBandit2019} discusses the process of productizing multi-armed bandits by determining context, sanity checking, evaluating offline, adding potential actions and logs, constraining, and engaging iterative improvements. Another study \cite{broden2018ensemble} discusses ensembling the content based and collaborative filtering based recommendations, using multi-armed bandits. The importance of exploration facing uncertainty is discussed by \cite{mcinerney2018explore}, and epsilon-greedy framework that learns explanation, content, and balance between exploration and exploitation jointly is provided as a solution. In addition, contextual bandits have been highlighted in \cite{tang2014ensemble} to resolve the cold-start problem. Authors from Yahoo! and Microsoft have extensively studied contextual bandit for news personalization \cite{chapelle2011empirical,mao2018batched,li2010contextual,li2012unbiased,agarwal2014taming,ross2013normalized}. Authors from Amazon also have highlighted the use of contextual bandit for optimizing content on web pages \cite{hill2017efficient}.  Authors from Google have long been arguing to use multi-armed bandit in massive online experimentation \cite{scott2015multi}. 

Furthermore, top tier tech players have presented research on other practical methods of recommendation algorithm evaluation, including the process analysis \cite{gunawardana2015evaluating}, the A/B testing \cite{gomez2016netflix,eide2018deep}, the offline-evaluation refinement \cite{rendle2019difficulty}, and an unbiased offline evaluation \cite{gruson2019offline,peska2018off,isinkaye2015recommendation}. Although, these are great studies which illuminate our way to build an infrastructure to leverage multi-armed bandit (MAB), none of them have mentioned their MAB solution as a production ready system to rapidly test various recommender system algorithms in an unbiased fashion. The current study aims to fill this gap. 

\section{Current and Selected Recommendation Algorithms}
Recommender systems can be classified by their input data, either the attributes or features of each item (e.g. price, rating, attributes) or user behavior (e.g. co-view within session, view-purchase within session, or historical purchase and views). The recommender systems that use the item attributes are called content based recommender system. The recommender systems that use the user behavior path can take various names based on their approach, including: collaborative filtering \cite{ekstrand2011collaborative}, matrix factorization \cite{johnson2014logistic,mnih2008probabilistic}, session based embedding recommender system \cite{wang2019survey,barkan2016item2vec,grbovic2018real,wang2018billion,arora2016simple,wang2019survey,pennington2014glove,caselles2018word2vec,wu2018session,bogina2017incorporating}, or probabilistic models \cite{smith2017two,chen2015optimal,wang2016knowledge,guigoures2018hierarchical,rendle2010factorizing}. There are hybrid recommender systems that combine both of the features and typically leverage deep neural networks \cite{cheng2016wide,wan2015next}, or factorization machines \cite{vasile2016meta,cheng2016wide,liang2016factorization}. 

The arms of a MAB model are the recommender system's variants. In this MAB campaign, we selected four different variants of recommender systems to test their performance online. We describe these four variants as follows:

\begin{itemize}
\item \textit{MAB \textunderscore ARM-1} includes a variant of content based recommender system which is built on a set of top of key item features.
\item \textit{MAB \textunderscore ARM-2} includes a variant of probabilistic recommender system. It is based on the conditional probability that a user clicks on item \textit{j} given they clicked on item \textit{i} within their shopping session.
\item \textit{MAB \textunderscore ARM-3} includes a variant of session based embedding recommender systems \cite{Oren2017Item2Vec}. It leverages user session activity data and estimates the similarity between the items in the embedding space. 
\item \textit{MAB \textunderscore ARM-4} includes a variant of the Matrix Factorization model \cite{liang2016factorization}. It creates a low dimensional representation for the items by using session co-view data.
\end{itemize}

We evaluate the online test results using two metrics - click-through-rate (CTR) and conversion-rate (CVR). CTR is measured as the proportion of users who clicked on recommendation items viewed. CVR is measured as the proportion of users who purchased on recommended items viewed. The former metric describes the user's engagement on the platform, while the later determines the conversion impact of the model. In the current study, we report both CTR and CVR online results and compare them across models. The online A/B test results is our gold standard. 

We describe the winner armed of previous MAB campaign and the control recommender system (i.e. status quo) as follows:
\begin{itemize}
    \item \textit{Control} variant includes current state-of-the-art model which is in production and generates recommendations to all live traffic in our platform. It leverages a neural-network architecture to create low dimensional embeddings for the items using user session clicks data similar to \cite{mitsoulis2019simple, grbovic2018real}.
    \item \textit{Previous campaign winner} variant includes a knowledge gradient model (e.g. \cite{chen2015optimal, wang2016knowledge}) based on beta binomial distribution. This model was a winner of the previous MAB campaign by earning the highest CTR among all the other models in that campaign.
\end{itemize}

\section{Multi-armed Bandit Architecture and Process}
Our daily mini-batch MAB training pipeline consists of three main processes: reward attribution, traffic proportion mini-batch process, and a randomized online traffic allocation. Figure 2 depicts the architecture of this pipeline.  
\begin{figure}[ht]
\centering
\includegraphics[width=\linewidth]{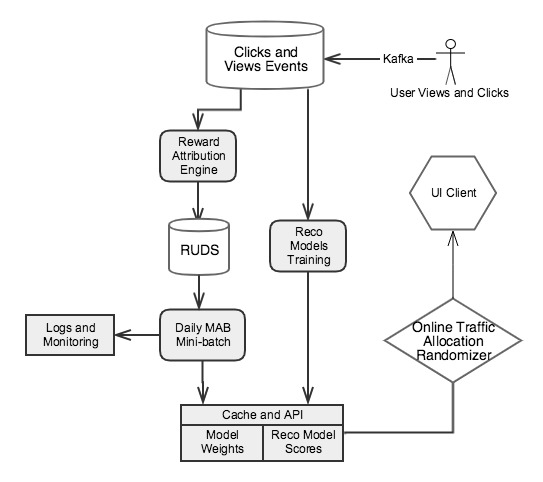}
\caption{Architecture diagram of MAB daily processing.}
\end{figure}
\subsection{Reward Attribution Engine}
The reward attribution layer leverages a real-time Kafka Queue \cite{kreps2011kafka} that receives messages from user interaction events, and back-end recommender service event logs, generating a Recommendation Unified Data set (RUDS). RUDS uses a configured look ahead window to join back-end recommendation service and front-end click and view events together. RUDS not only captures the micro-conversion events (e.g. clicks and views), but also the ultimate conversion event (e.g. purchase), allowing near real-time performance monitoring of each recommender system variant. We leverage a visitor based metric, as opposed to visit based metric, to minimize the potential statistical dependency between the observations. 

\subsection{MAB Daily Mini-batch}
This layer receives sufficient statistics (i.e. number of views and clicks for each variant) from the reward attribution engine and pushes the computed traffic proportions for each armed to the cache. For this traffic allocation, we leverage Thompson sampling (TS). Since the sufficient statistics can be modeled as Bernoulli random variables with parameter ($p$ = probably of a click), it is straightforward to assume ($p$) follows a Beta distribution. This formulation allows us to leverage TS without the need for numerical approximation since Beta is a conjugate prior. As visitors click on recommended items each day, we update our posterior distributions which also has Beta distribution. This process is formalized as follows \cite{shipra2012thompson}. It initially assumes armed i to have a uniform prior $Beta(1,1)$ for probability of success (user click) $p_i$. At epoch (e.g. day) $t$, having observed $S_i(t)$ accumulated successes and $F_i(t)$ accumulated failures from the first epoch, the algorithm updates the posterior distribution of $p_i$ to $Beta(S_i(t)+1, F_i(t)+1)$. The algorithm then samples from these posterior distributions of $p_i$'s and allocates the traffic to the arms according to the proportion of simulated samples that $p_i$ dominates all other arms. In our case, we draw $10,000$ samples from this posterior distribution at each mini-batch daily run. For each sample we compare the drawn CTR across arms, and we identify the armed that has maximum CTR. Then, we compute the proportion that each armed had maximum drawn CTR out of all the draws. The computed traffic allocation is then pushed to a cache. This traffic allocation data in the cache is exposed by an API is then leveraged by the online traffic allocation randomizer for the next day, which we describe next. In addition, we log these intermediary spin offs and the final traffic allocations for monitoring purposes. The whole MAB process has low memory requirement and constant time complexity.

\subsection{Online Traffic Allocation Randomizer}
When a visitor visits our website on a item page, a request is sent to the recommendation service. This Java-based process is the orchestrator of variants of the recommender systems. This process calls the traffic allocation proportion cache and draws a uniform random variable between $0$ and $1$ for each visitor. Based on the bucket of the cumulative distribution of allocation proportion that this random variable instance falls into, the recommender service calls the relevant recommender system variant and returns the list of recommended items. Then, this process sends the recommended item to UI client that is equipped with tracking script to show them to the users.

\subsection{Building a Robust System}
We conclude this section by presenting a couple of practical approaches we leveraged to make the MAB pipeline robust. First, our business is seasonal and subject to daily, weekly, and monthly effects. The risk of using MAB in this context is that, a variant might perform poorly at a given date, while being globally optimal. Another risk could be that there is no globally optimal variant. In this case, for each time period the optimal variant may be different, violating the stationarity assumption. In this case, the vanilla bandit might allocate zero traffic at the first mini-batch run. This zero traffic allocation will take the opportunity from this optimal armed to win in future dates. To resolve this issue we defined a lower bound on the traffic allocation, so if the traffic allocation of an armed is less than the configurable secured traffic threshold, we take traffic from winner armed proportional to their optimality and allocate to the loser arm. Changing this lower bound adaptively according to a configurable schedule allows us to give enough opportunity to the arms that might lose at initial days of the campaign, but might win afterward. In practice, we have observed that variants that lose the initial days can win back traffic due to this protective process that we have embedded in the pipeline. Second, our platform is subject to many web crawler and scraper bots, which might skew the traffic allocation. To solve this issue, in addition to leveraging visitor based click metric, we also leveraged mini-batch approach. In other words, our traffic proportion computation process is run in nightly batches, to aggregate data, and not be sensitive to noise. Indeed, this aggregation reduces unbiased noise, so we rely on our bot detection service to filter out the bot biased noise. 

Third, although we run the MAB as a campaign as opposed to restless bandit, we designed our pipeline so that in case we add a new variant mid-way, it starts with uniform prior, and the traffic is re-allocated as we collect more data at nightly batches. Fourth, it is possible that a given recommender system variant has defects, but we don't want to stop all the variants, only because one of the variants has defect. To accommodate this requirement, we defined a black listing process, which allows us to put a given variant into a black list to zero out the traffic allocated to it the next day. Fifth, our MAB pipeline should be robust to any pipeline breakage in upstream system. To achieve robustness, we modified our mini-batch query from RUDS to not change the traffic allocation, when such a breakage happens. This in-variance under no data condition is consistent with Bayesian principles.

\section{Experimental Methods}

In this section we describe the two components of our framework, offline metrics and online MAB testing. Furthermore, we describe the benefits of our MAB approach. 

\subsection{Offline Evaluation}
Building offline metrics for recommendation models is a tricky task, and it can vary by use-case and industry. We approach the recommendation task as a ranking problem, meaning that we are interested in showing relatively few items to the user that we consider most relevant. Accordingly, we use popular ranking metrics - Normalized Discounted Cumulative Gain (NDCG), Mean Reciprocal Rank (MRR), and Mean Average Precision(MAP) to evaluate the offline performance of the models \cite{le2007direct, sirotkin2013search}. In most cases, the offline evaluation has limitations due to the bias in the underlying data \cite{beel2013comparative}. This selection bias might stem from current system constraints (e.g. state of user interface) and search cost, which lead the user to choose the items under bounded rationality \cite{gabaix2006costly,simon1972theories}.


We sampled multiple months of the users' previous logged interaction data, containing clicks and purchases. We segmented this into two interaction matrices- one based on clicks and the other based on purchases or bookings. We computed offline metrics for these models on both interaction matrices and used this data as a source to compare the offline performance of all the recommender system models in this MAB campaign.



\begin{table*}[ht]
    \caption{Offline and online performance of all selected recommender system models in this MAB campaign}
    \begin{threeparttable}
    \begin{center}
    \begin{tabular}{|c|c c c|c c c|c|c c|}
    \hline
      & 
        \multicolumn{6}{|c|}{Offline results} & \multicolumn{3}{|c|}{online results}\\
        \cline{2-10}
        Recommender Systems & \multicolumn{3}{|c|}{CTR} & \multicolumn{3}{|c|}{CVR}
        & MAB test & \multicolumn{2}{|c|}{A/B test} \\
        \cline{2-7}
        &  MRR & NDCG@10 & MAP@10 & MRR & NDCG@10 & MAP@10 & *CTR(\%) & *CTR(\%) & *CVR(\%)\\
    \hline 
    MAB\textunderscore ARM-1 & 0.206 & 0.338 & 0.180 & 0.232 & 0.427 & 0.124 & -0.25 & - & -\\
    MAB\textunderscore ARM-2 & \textbf{0.585} & \textbf{0.800} & \textbf{0.595} & 0.256 & 0.441 & 0.146 & +11.16 & - & -\\
    MAB\textunderscore ARM-3 & 0.336 & 0.582 & 0.390 & 0.299 & 0.496 & 0.167 & +0.45 & - & - \\ 
    MAB\textunderscore ARM-4 & 0.430 & 0.672 & 0.470 & \textbf{0.491} & \textbf{0.670} & \textbf{0.292} & -0.15 & - & -\\ 
    previous campaign winner & \textbf{0.585} & \textbf{0.800} & 0.592 & 0.256 & 0.442 & 0.146 & - & $+10.8\pm0.43$ & $+0.3\pm 0.87$ \\
    Control & 0.291 & 0.505 & 0.349 & 0.302 & 0.504 & 0.18 & - & - & -\\
    \hline
    \end{tabular}
    \begin{tablenotes}\footnotesize
    \item * percentage difference in comparison to control model.
    \end{tablenotes}
    \end{center}
    \end{threeparttable}
    \label{tab:table1}
\end{table*}

\subsection{Online MAB Testing}

\begin{figure}[ht]
\centering
\includegraphics[width=\linewidth]{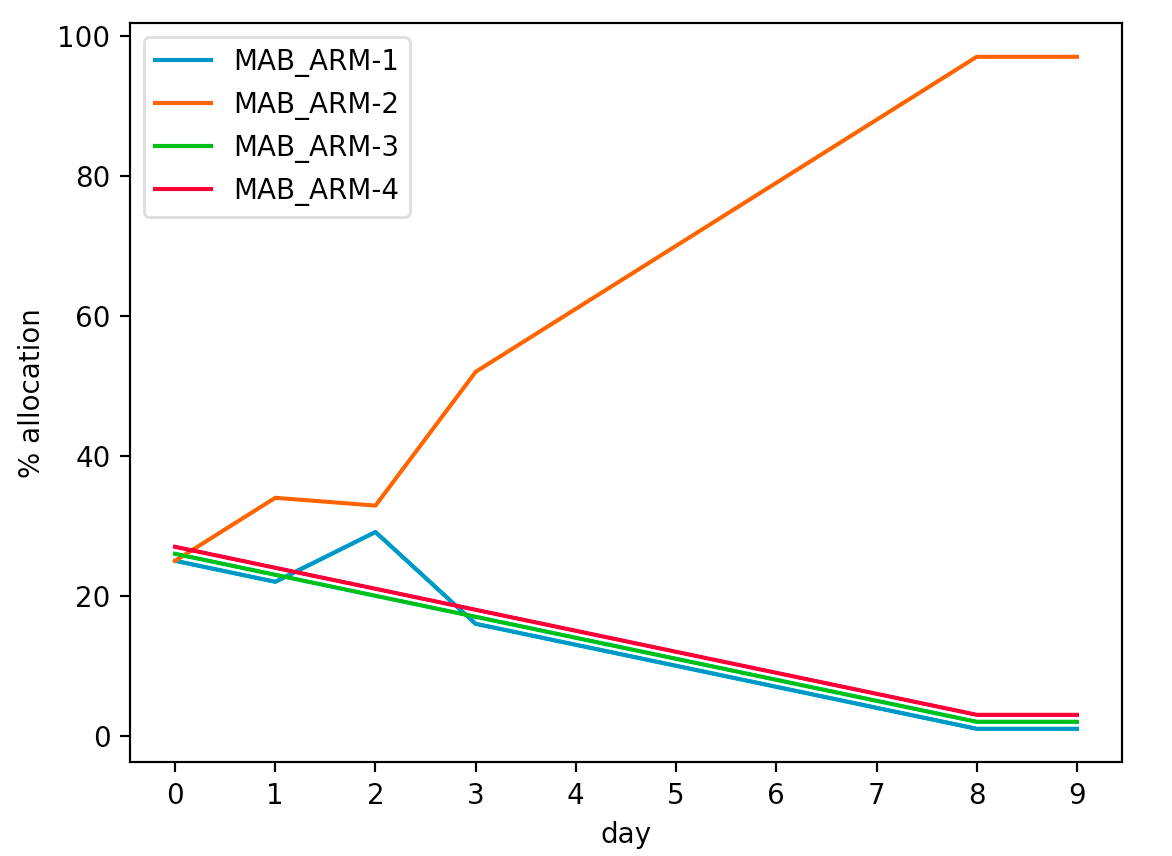}
\caption{Multi-armed Bandit Re-Allocation Timeseries}
\label{fig:trafficAlloc}
\end{figure}

In every MAB campaign, we test the performance of new variants of recommender system models based on CTR. Once the campaign is ended, we will A/B test the winner armed of the MAB model against current control to measure it's impact on both CTR and CVR. In this way, we keep track of conversion besides click through rate. The MAB setting is as follows: at each round we have a set of actions $A$. After choosing an action $a \in A$, we observe a reward $r$. The goal is to find a policy that selects actions such that cumulative reward across epochs is as large as possible. Equivalently, such policy minimizes regret relative to the best action policy, known ex-post. In our case, since the reward is either $0$ or $1$ (whether the user clicks on the viewed recommended item or not), it can be treated as a Bernoulli bandit problem. 

To solve this MAB problem, there are several algorithms to select from including: Thompson sampling (TS), Upper Confidence Bounds (UCB), and Epsilon-Greedy (EG). We used TS due to its optimality and robustness to noise in the production environment, since it leverages a Bayesian framework \cite{broden2018ensemble}. These noises can stem from upstream system and data dependencies. In addition, in contrast to UCB, TS will allow us to do randomization at each mini-batch round, which prevents the risk of falling into local optima. 



\subsection{Benefits of MAB approach}
Our online A/B testing platform is stateful, meaning for any test, it randomly buckets users to variants of the test and once a user is bucketed, they will remain in the same variant for the whole test. This is to keep the user's experience consistent on our platform and also to measure the performance of variants in a user independent fashion. However, for recommender systems, the consistency of experiment is not required, meaning it doesn't hurt if a user sees items $l_1$, $l_2$, $l_3$, ... as recommendations for an item $l_0$ today, and sees items $l_3$, $l_5$, $l_7$, .. as the recommendations for the same item next day. Indeed such an approach is helpful to create diversity. In addition, even if we don't do this, it is possible that few of recommended items rendered to a user might become unavailable over time, and as a result the new list of recommendations are different. Furthermore, we gain statistical strength in MAB by increasing traffic for the winner armed and decreasing it for the arms that are not winning. As a result, we give it more exposure and traffic this way.

\section{Results}

Table \ref{tab:table1} displays the results of all three components of the framework.  In the offline metric section (left),  \textit{MAB\textunderscore ARM-2} performs the best across all CTR offline metrics compared to all the other arms in the current campaign. Its offline performance based on CTR is very close to the previous campaign winner and is superior to control. In contrast, \textit{MAB\textunderscore ARM-4} has better CVR offline metrics compared to all the other models.  Also, the offline performance of previous campaign winner model is superior to control based on CTR but is inferior based on CVR.

Figure \ref{fig:trafficAlloc} shows the results of online traffic allocation for the four variants of this campaign. In MAB test portion of Table \ref{tab:table1}, we observe that MAB\textunderscore ARM-2 has the highest expected CTR followed by MAB\textunderscore ARM-3, MAB\textunderscore ARM-4, and MAB\textunderscore ARM-1. We started this MAB campaign with equal traffic allocation for all the arms. It is clear that our algorithm incrementally shifted the traffic from other arms to MAB\textunderscore ARM-2 and within a two weeks it allocated all the traffic to that arm. This result was expected, as from our offline evaluation results we observed MAB\textunderscore ARM-2 has the highest best online CTR metrics.  

We performed a traditional A/B test to measure the engagement and conversion impact of the \textit{previous winner arm} against the \textit{Control} model. In our A/B test, we observed an incremental increase in both $10.8\%$ CTR. This result was expected, as it has higher NDCG for CTR compared to control from our offline evaluation.  Also, in our A/B test we observed incremental $0.3\%$ CVR but it was not statistical significant. This result was not expected because it outperformed the control model despite it's lower offline CVR metrics

\section{Discussion}

For a long time, our team has been aware that CTR does not necessarily predict CVR. A particular algorithm may present items that engage users' curiosity but are not necessarily relevant for their needs. For instance, a pretty or novel picture might invite people to click on an item, especially if the users are at an exploratory phase.  However, that interaction may have also been a lost opportunity to show a truly relevant or personalized item. Therefore, an increase in CTR could actually hurt CVR.  Likewise, an algorithm could improve the relevance of items at a level that is only apparent after one has clicked through to see the item, thereby increasing CVR while not affecting CTR.  Ideally, an algorithm would improve both metrics if, by design, it is presenting more relevant items at the outset.  Nevertheless, our framework evaluates both metrics at every step. 

Our current case study presents a good example of how one cannot use any one method, offline or online, to evaluate algorithms.  According to the offline metrics, \textit{MAB\textunderscore ARM-2} and the previous campaign winner should be clearly superior on CTR and mediocre on CVR.  On the other hand, \textit{MAB\textunderscore ARM-4} is clearly superior across all CVR metrics while being the 2nd best armed in terms of CTR.  Therefore both \textit{MAB\textunderscore ARM-4} and \textit{MAB\textunderscore ARM-2} may be better than the previous winner and also control, but for different reasons.  However, these results were not exactly consistent with the MAB test. 

The purpose of the MAB-test is to give additional information to select candidate algorithms for A/B test in the next campaign.  Unsurprisingly, \textit{MAB\textunderscore ARM-2} quickly gained allocation to due its higher CTR. So, it is arguably the most viable candidate.  Despite losing on CTR, \textit{MAB\textunderscore ARM-4} still could perform well on CVR during the next A/B test. So, it will go into the next campaign as well.  Since \textit{MAB\textunderscore ARM-1} and \textit{MAB\textunderscore ARM-3} did not perform well on online CTR, they will not be included.  But, it is worthwhile to note that the offline CTR metrics would suggest that the \textit{MAB\textunderscore ARM-4} would outperform \textit{MAB\textunderscore ARM-3} on CTR, which was not the case. 

The A/B test within which the MAB test was conducted succeeded in validating that the previous campaign winner does beat control on both CTR and CVR.  This is supported by the offline metrics for CTR but contradicts the CVR metrics.  The difference between the offline metrics were small between these two algorithms, suggesting small differences may be within the noise. 

\section{Conclusion}

In this paper we discussed how we productionalized MAB to evaluate recommender system variants in a less biased fashion. We discussed how MAB can help us save time in evaluating recommender system variants in our two-sided short term vacation rental marketplace. In addition, we mentioned various practical approaches we leveraged to make MAB robust when the business is seasonal and data is non-stationary. We presented the result of one of the MAB campaigns we ran, and its experimental outcomes. Our results allowed us to compare A/B tests and MAB tests across a small number of variants across all entities. Such a comparison is not possible in the scenarios in which MAB is implemented as one bandit per entity.

There are several open issues to be addressed in future work. In particular, we are going to scale our MAB framework to treat each item in our platform as a MAB problem. Furthermore, we are planning to extend our framework to enable multi-objective optimization. 

\begin{acks}
The authors would like to thank Travis Brady, Pavlos Mitsoulis Ntompos, Ben Dundee, Kurt Smith, and John Meakin for their internal review of this paper and their helpful feedback.

\end{acks}

\bibliographystyle{ACM-Reference-Format}
\bibliography{mab_paper}


\begin{thebibliography}{86}


\ifx \showCODEN    \undefined \def \showCODEN     #1{\unskip}     \fi
\ifx \showDOI      \undefined \def \showDOI       #1{#1}\fi
\ifx \showISBNx    \undefined \def \showISBNx     #1{\unskip}     \fi
\ifx \showISBNxiii \undefined \def \showISBNxiii  #1{\unskip}     \fi
\ifx \showISSN     \undefined \def \showISSN      #1{\unskip}     \fi
\ifx \showLCCN     \undefined \def \showLCCN      #1{\unskip}     \fi
\ifx \shownote     \undefined \def \shownote      #1{#1}          \fi
\ifx \showarticletitle \undefined \def \showarticletitle #1{#1}   \fi
\ifx \showURL      \undefined \def \showURL       {\relax}        \fi
\providecommand\bibfield[2]{#2}
\providecommand\bibinfo[2]{#2}
\providecommand\natexlab[1]{#1}
\providecommand\showeprint[2][]{arXiv:#2}

\bibitem[\protect\citeauthoryear{Abensur and Orlov}{Abensur and Orlov}{2019}]%
        {ProductizationContextualBandit2019}
\bibfield{author}{\bibinfo{person}{Balashov I. Bar S. Lempel R. Moscovici~N.
  Abensur, D.} {and} \bibinfo{person}{I. Orlov}.}
  \bibinfo{year}{2019}\natexlab{}.
\newblock \showarticletitle{{Productization Challenges of Contextual
  Multi-Armed Bandits}}.
\newblock  (\bibinfo{year}{2019}).
\newblock


\bibitem[\protect\citeauthoryear{Agarwal, Hsu, Kale, Langford, Li, and
  Schapire}{Agarwal et~al\mbox{.}}{2014}]%
        {agarwal2014taming}
\bibfield{author}{\bibinfo{person}{Alekh Agarwal}, \bibinfo{person}{Daniel
  Hsu}, \bibinfo{person}{Satyen Kale}, \bibinfo{person}{John Langford},
  \bibinfo{person}{Lihong Li}, {and} \bibinfo{person}{Robert Schapire}.}
  \bibinfo{year}{2014}\natexlab{}.
\newblock \showarticletitle{Taming the monster: A fast and simple algorithm for
  contextual bandits}. In \bibinfo{booktitle}{\emph{International Conference on
  Machine Learning}}. \bibinfo{pages}{1638--1646}.
\newblock


\bibitem[\protect\citeauthoryear{Agarwal, Luo, Neyshabur, and Schapire}{Agarwal
  et~al\mbox{.}}{2016}]%
        {agarwal2016corralling}
\bibfield{author}{\bibinfo{person}{Alekh Agarwal}, \bibinfo{person}{Haipeng
  Luo}, \bibinfo{person}{Behnam Neyshabur}, {and} \bibinfo{person}{Robert~E
  Schapire}.} \bibinfo{year}{2016}\natexlab{}.
\newblock \showarticletitle{Corralling a band of bandit algorithms}.
\newblock \bibinfo{journal}{\emph{arXiv preprint arXiv:1612.06246}}
  (\bibinfo{year}{2016}).
\newblock


\bibitem[\protect\citeauthoryear{Ahonen et~al\mbox{.}}{Ahonen
  et~al\mbox{.}}{2017}]%
        {ahonen2017applying}
\bibfield{author}{\bibinfo{person}{Niko-Petteri Ahonen} {et~al\mbox{.}}}
  \bibinfo{year}{2017}\natexlab{}.
\newblock \showarticletitle{Applying Bayesian Bandits For Solving Optimal
  Budget Allocation In Social Media Marketing}.
\newblock  (\bibinfo{year}{2017}).
\newblock


\bibitem[\protect\citeauthoryear{Arora, Liang, and Ma}{Arora
  et~al\mbox{.}}{2016}]%
        {arora2016simple}
\bibfield{author}{\bibinfo{person}{Sanjeev Arora}, \bibinfo{person}{Yingyu
  Liang}, {and} \bibinfo{person}{Tengyu Ma}.} \bibinfo{year}{2016}\natexlab{}.
\newblock \showarticletitle{A simple but tough-to-beat baseline for sentence
  embeddings}.
\newblock  (\bibinfo{year}{2016}).
\newblock


\bibitem[\protect\citeauthoryear{Audibert and Bubeck}{Audibert and
  Bubeck}{2010}]%
        {audibert2010regret}
\bibfield{author}{\bibinfo{person}{Jean-Yves Audibert} {and}
  \bibinfo{person}{S{e}bastien Bubeck}.} \bibinfo{year}{2010}\natexlab{}.
\newblock \showarticletitle{Regret bounds and minimax policies under partial
  monitoring}.
\newblock \bibinfo{journal}{\emph{Journal of Machine Learning Research}}
  \bibinfo{volume}{11}, \bibinfo{number}{Oct} (\bibinfo{year}{2010}),
  \bibinfo{pages}{2785--2836}.
\newblock


\bibitem[\protect\citeauthoryear{Avner and Mannor}{Avner and Mannor}{2014}]%
        {avner2014concurrent}
\bibfield{author}{\bibinfo{person}{Orly Avner} {and} \bibinfo{person}{Shie
  Mannor}.} \bibinfo{year}{2014}\natexlab{}.
\newblock \showarticletitle{Concurrent bandits and cognitive radio networks}.
  In \bibinfo{booktitle}{\emph{Joint European Conference on Machine Learning
  and Knowledge Discovery in Databases}}. Springer, \bibinfo{pages}{66--81}.
\newblock


\bibitem[\protect\citeauthoryear{Barkan and Koenigstein}{Barkan and
  Koenigstein}{2016}]%
        {barkan2016item2vec}
\bibfield{author}{\bibinfo{person}{Oren Barkan} {and} \bibinfo{person}{Noam
  Koenigstein}.} \bibinfo{year}{2016}\natexlab{}.
\newblock \showarticletitle{Item2vec: neural item embedding for collaborative
  filtering}. In \bibinfo{booktitle}{\emph{2016 IEEE 26th International
  Workshop on Machine Learning for Signal Processing (MLSP)}}. IEEE,
  \bibinfo{pages}{1--6}.
\newblock


\bibitem[\protect\citeauthoryear{Beel, Genzmehr, Langer, N{\"u}rnberger, and
  Gipp}{Beel et~al\mbox{.}}{2013}]%
        {beel2013comparative}
\bibfield{author}{\bibinfo{person}{Joeran Beel}, \bibinfo{person}{Marcel
  Genzmehr}, \bibinfo{person}{Stefan Langer}, \bibinfo{person}{Andreas
  N{\"u}rnberger}, {and} \bibinfo{person}{Bela Gipp}.}
  \bibinfo{year}{2013}\natexlab{}.
\newblock \showarticletitle{A comparative analysis of offline and online
  evaluations and discussion of research paper recommender system evaluation}.
  In \bibinfo{booktitle}{\emph{Proceedings of the international workshop on
  reproducibility and replication in recommender systems evaluation}}. ACM,
  \bibinfo{pages}{7--14}.
\newblock


\bibitem[\protect\citeauthoryear{Besson and Kaufmann}{Besson and
  Kaufmann}{2017}]%
        {besson2017multi}
\bibfield{author}{\bibinfo{person}{Lilian Besson} {and} \bibinfo{person}{Emilie
  Kaufmann}.} \bibinfo{year}{2017}\natexlab{}.
\newblock \showarticletitle{Multi-player bandits revisited}.
\newblock \bibinfo{journal}{\emph{arXiv preprint arXiv:1711.02317}}
  (\bibinfo{year}{2017}).
\newblock


\bibitem[\protect\citeauthoryear{Bogina and Kuflik}{Bogina and Kuflik}{2017}]%
        {bogina2017incorporating}
\bibfield{author}{\bibinfo{person}{Veronika Bogina} {and} \bibinfo{person}{Tsvi
  Kuflik}.} \bibinfo{year}{2017}\natexlab{}.
\newblock \showarticletitle{Incorporating Dwell Time in Session-Based
  Recommendations with Recurrent Neural Networks.}. In
  \bibinfo{booktitle}{\emph{RecTemp@ RecSys}}. \bibinfo{pages}{57--59}.
\newblock


\bibitem[\protect\citeauthoryear{Boursier and Perchet}{Boursier and
  Perchet}{2018}]%
        {boursier2018sic}
\bibfield{author}{\bibinfo{person}{Etienne Boursier} {and}
  \bibinfo{person}{Vianney Perchet}.} \bibinfo{year}{2018}\natexlab{}.
\newblock \showarticletitle{SIC-MMAB: synchronisation involves communication in
  multiplayer multi-armed bandits}.
\newblock \bibinfo{journal}{\emph{arXiv preprint arXiv:1809.08151}}
  (\bibinfo{year}{2018}).
\newblock


\bibitem[\protect\citeauthoryear{Brod{\'e}n, Hammar, Nilsson, and
  Paraschakis}{Brod{\'e}n et~al\mbox{.}}{2018}]%
        {broden2018ensemble}
\bibfield{author}{\bibinfo{person}{Bj{\"o}rn Brod{\'e}n},
  \bibinfo{person}{Mikael Hammar}, \bibinfo{person}{Bengt~J Nilsson}, {and}
  \bibinfo{person}{Dimitris Paraschakis}.} \bibinfo{year}{2018}\natexlab{}.
\newblock \showarticletitle{Ensemble recommendations via Thompson sampling: an
  experimental study within e-Commerce}. In \bibinfo{booktitle}{\emph{23rd
  International Conference on Intelligent User Interfaces}}. ACM,
  \bibinfo{pages}{19--29}.
\newblock


\bibitem[\protect\citeauthoryear{Bubeck, Cesa-Bianchi, et~al\mbox{.}}{Bubeck
  et~al\mbox{.}}{2012}]%
        {bubeck2012regret}
\bibfield{author}{\bibinfo{person}{S{\'e}bastien Bubeck},
  \bibinfo{person}{Nicolo Cesa-Bianchi}, {et~al\mbox{.}}}
  \bibinfo{year}{2012}\natexlab{}.
\newblock \showarticletitle{Regret analysis of stochastic and nonstochastic
  multi-armed bandit problems}.
\newblock \bibinfo{journal}{\emph{Foundations and Trends{\textregistered} in
  Machine Learning}} \bibinfo{volume}{5}, \bibinfo{number}{1}
  (\bibinfo{year}{2012}), \bibinfo{pages}{1--122}.
\newblock


\bibitem[\protect\citeauthoryear{Cao, Wen, Kveton, and Xie}{Cao
  et~al\mbox{.}}{2018}]%
        {cao2018nearly}
\bibfield{author}{\bibinfo{person}{Yang Cao}, \bibinfo{person}{Zheng Wen},
  \bibinfo{person}{Branislav Kveton}, {and} \bibinfo{person}{Yao Xie}.}
  \bibinfo{year}{2018}\natexlab{}.
\newblock \showarticletitle{Nearly Optimal Adaptive Procedure with Change
  Detection for Piecewise-Stationary Bandit}.
\newblock \bibinfo{journal}{\emph{arXiv preprint arXiv:1802.03692}}
  (\bibinfo{year}{2018}).
\newblock


\bibitem[\protect\citeauthoryear{Capp{\'e}, Garivier, Maillard, Munos, Stoltz,
  et~al\mbox{.}}{Capp{\'e} et~al\mbox{.}}{2013}]%
        {cappe2013kullback}
\bibfield{author}{\bibinfo{person}{Olivier Capp{\'e}},
  \bibinfo{person}{Aur{\'e}lien Garivier}, \bibinfo{person}{Odalric-Ambrym
  Maillard}, \bibinfo{person}{R{\'e}mi Munos}, \bibinfo{person}{Gilles Stoltz},
  {et~al\mbox{.}}} \bibinfo{year}{2013}\natexlab{}.
\newblock \showarticletitle{Kullback--leibler upper confidence bounds for
  optimal sequential allocation}.
\newblock \bibinfo{journal}{\emph{The Annals of Statistics}}
  \bibinfo{volume}{41}, \bibinfo{number}{3} (\bibinfo{year}{2013}),
  \bibinfo{pages}{1516--1541}.
\newblock


\bibitem[\protect\citeauthoryear{Caselles-Dupr{\'e}, Lesaint, and
  Royo-Letelier}{Caselles-Dupr{\'e} et~al\mbox{.}}{2018}]%
        {caselles2018word2vec}
\bibfield{author}{\bibinfo{person}{Hugo Caselles-Dupr{\'e}},
  \bibinfo{person}{Florian Lesaint}, {and} \bibinfo{person}{Jimena
  Royo-Letelier}.} \bibinfo{year}{2018}\natexlab{}.
\newblock \showarticletitle{Word2vec applied to recommendation: Hyperparameters
  matter}. In \bibinfo{booktitle}{\emph{Proceedings of the 12th ACM Conference
  on Recommender Systems}}. ACM, \bibinfo{pages}{352--356}.
\newblock


\bibitem[\protect\citeauthoryear{Cesa-Bianchi, Gentile, Lugosi, and
  Neu}{Cesa-Bianchi et~al\mbox{.}}{2017}]%
        {cesa2017boltzmann}
\bibfield{author}{\bibinfo{person}{Nicol{\`o} Cesa-Bianchi},
  \bibinfo{person}{Claudio Gentile}, \bibinfo{person}{G{\'a}bor Lugosi}, {and}
  \bibinfo{person}{Gergely Neu}.} \bibinfo{year}{2017}\natexlab{}.
\newblock \showarticletitle{Boltzmann exploration done right}. In
  \bibinfo{booktitle}{\emph{Advances in Neural Information Processing
  Systems}}. \bibinfo{pages}{6284--6293}.
\newblock


\bibitem[\protect\citeauthoryear{Chapelle and Li}{Chapelle and Li}{2011}]%
        {chapelle2011empirical}
\bibfield{author}{\bibinfo{person}{Olivier Chapelle} {and}
  \bibinfo{person}{Lihong Li}.} \bibinfo{year}{2011}\natexlab{}.
\newblock \showarticletitle{An empirical evaluation of thompson sampling}. In
  \bibinfo{booktitle}{\emph{Advances in neural information processing
  systems}}. \bibinfo{pages}{2249--2257}.
\newblock


\bibitem[\protect\citeauthoryear{Chen, Reyes, Gupta, McAlpine, and Powell}{Chen
  et~al\mbox{.}}{2015}]%
        {chen2015optimal}
\bibfield{author}{\bibinfo{person}{Si Chen}, \bibinfo{person}{Kristofer-Roy~G
  Reyes}, \bibinfo{person}{Maneesh~K Gupta}, \bibinfo{person}{Michael~C
  McAlpine}, {and} \bibinfo{person}{Warren~B Powell}.}
  \bibinfo{year}{2015}\natexlab{}.
\newblock \showarticletitle{Optimal learning in experimental design using the
  knowledge gradient policy with application to characterizing nanoemulsion
  stability}.
\newblock \bibinfo{journal}{\emph{SIAM/ASA Journal on Uncertainty
  Quantification}} \bibinfo{volume}{3}, \bibinfo{number}{1}
  (\bibinfo{year}{2015}), \bibinfo{pages}{320--345}.
\newblock


\bibitem[\protect\citeauthoryear{Cheng, Koc, Harmsen, Shaked, Chandra, Aradhye,
  Anderson, Corrado, Chai, Ispir, et~al\mbox{.}}{Cheng et~al\mbox{.}}{2016}]%
        {cheng2016wide}
\bibfield{author}{\bibinfo{person}{Heng-Tze Cheng}, \bibinfo{person}{Levent
  Koc}, \bibinfo{person}{Jeremiah Harmsen}, \bibinfo{person}{Tal Shaked},
  \bibinfo{person}{Tushar Chandra}, \bibinfo{person}{Hrishi Aradhye},
  \bibinfo{person}{Glen Anderson}, \bibinfo{person}{Greg Corrado},
  \bibinfo{person}{Wei Chai}, \bibinfo{person}{Mustafa Ispir}, {et~al\mbox{.}}}
  \bibinfo{year}{2016}\natexlab{}.
\newblock \showarticletitle{Wide \& deep learning for recommender systems}. In
  \bibinfo{booktitle}{\emph{Proceedings of the 1st workshop on deep learning
  for recommender systems}}. ACM, \bibinfo{pages}{7--10}.
\newblock


\bibitem[\protect\citeauthoryear{Combes, Magureanu, and Proutiere}{Combes
  et~al\mbox{.}}{2017}]%
        {combes2017minimal}
\bibfield{author}{\bibinfo{person}{Richard Combes}, \bibinfo{person}{Stefan
  Magureanu}, {and} \bibinfo{person}{Alexandre Proutiere}.}
  \bibinfo{year}{2017}\natexlab{}.
\newblock \showarticletitle{Minimal exploration in structured stochastic
  bandits}. In \bibinfo{booktitle}{\emph{Advances in Neural Information
  Processing Systems}}. \bibinfo{pages}{1763--1771}.
\newblock


\bibitem[\protect\citeauthoryear{Eide and Zhou}{Eide and Zhou}{2018}]%
        {eide2018deep}
\bibfield{author}{\bibinfo{person}{Simen Eide} {and} \bibinfo{person}{Ning
  Zhou}.} \bibinfo{year}{2018}\natexlab{}.
\newblock \showarticletitle{Deep neural network marketplace recommenders in
  online experiments}. In \bibinfo{booktitle}{\emph{Proceedings of the 12th ACM
  Conference on Recommender Systems}}. ACM, \bibinfo{pages}{387--391}.
\newblock


\bibitem[\protect\citeauthoryear{Ekstrand, Riedl, Konstan,
  et~al\mbox{.}}{Ekstrand et~al\mbox{.}}{2011}]%
        {ekstrand2011collaborative}
\bibfield{author}{\bibinfo{person}{Michael~D Ekstrand}, \bibinfo{person}{John~T
  Riedl}, \bibinfo{person}{Joseph~A Konstan}, {et~al\mbox{.}}}
  \bibinfo{year}{2011}\natexlab{}.
\newblock \showarticletitle{Collaborative filtering recommender systems}.
\newblock \bibinfo{journal}{\emph{Foundations and Trends{\textregistered} in
  Human--Computer Interaction}} \bibinfo{volume}{4}, \bibinfo{number}{2}
  (\bibinfo{year}{2011}), \bibinfo{pages}{81--173}.
\newblock


\bibitem[\protect\citeauthoryear{Evirgen and Kose}{Evirgen and Kose}{2017}]%
        {evirgen2017effect}
\bibfield{author}{\bibinfo{person}{Noyan Evirgen} {and} \bibinfo{person}{Alper
  Kose}.} \bibinfo{year}{2017}\natexlab{}.
\newblock \showarticletitle{The effect of communication on noncooperative
  multiplayer multi-armed bandit problems}. In \bibinfo{booktitle}{\emph{2017
  16th IEEE International Conference on Machine Learning and Applications
  (ICMLA)}}. IEEE, \bibinfo{pages}{331--336}.
\newblock


\bibitem[\protect\citeauthoryear{Filippi, Capp{\'e}, and Garivier}{Filippi
  et~al\mbox{.}}{2010}]%
        {filippi2010optimism}
\bibfield{author}{\bibinfo{person}{Sarah Filippi}, \bibinfo{person}{Olivier
  Capp{\'e}}, {and} \bibinfo{person}{Aur{\'e}lien Garivier}.}
  \bibinfo{year}{2010}\natexlab{}.
\newblock \showarticletitle{Optimism in reinforcement learning and
  Kullback-Leibler divergence}. In \bibinfo{booktitle}{\emph{2010 48th Annual
  Allerton Conference on Communication, Control, and Computing (Allerton)}}.
  IEEE, \bibinfo{pages}{115--122}.
\newblock


\bibitem[\protect\citeauthoryear{Gabaix, Laibson, Moloche, and Weinberg}{Gabaix
  et~al\mbox{.}}{2006}]%
        {gabaix2006costly}
\bibfield{author}{\bibinfo{person}{Xavier Gabaix}, \bibinfo{person}{David
  Laibson}, \bibinfo{person}{Guillermo Moloche}, {and} \bibinfo{person}{Stephen
  Weinberg}.} \bibinfo{year}{2006}\natexlab{}.
\newblock \showarticletitle{Costly information acquisition: Experimental
  analysis of a boundedly rational model}.
\newblock \bibinfo{journal}{\emph{American Economic Review}}
  \bibinfo{volume}{96}, \bibinfo{number}{4} (\bibinfo{year}{2006}),
  \bibinfo{pages}{1043--1068}.
\newblock


\bibitem[\protect\citeauthoryear{Garivier and Capp{\'e}}{Garivier and
  Capp{\'e}}{2011}]%
        {garivier2011kl}
\bibfield{author}{\bibinfo{person}{Aur{\'e}lien Garivier} {and}
  \bibinfo{person}{Olivier Capp{\'e}}.} \bibinfo{year}{2011}\natexlab{}.
\newblock \showarticletitle{The KL-UCB algorithm for bounded stochastic bandits
  and beyond}. In \bibinfo{booktitle}{\emph{Proceedings of the 24th annual
  conference on learning theory}}. \bibinfo{pages}{359--376}.
\newblock


\bibitem[\protect\citeauthoryear{Garivier, Hadiji, Menard, and Stoltz}{Garivier
  et~al\mbox{.}}{2018}]%
        {garivier2018kl}
\bibfield{author}{\bibinfo{person}{Aur{\'e}lien Garivier},
  \bibinfo{person}{H{\'e}di Hadiji}, \bibinfo{person}{Pierre Menard}, {and}
  \bibinfo{person}{Gilles Stoltz}.} \bibinfo{year}{2018}\natexlab{}.
\newblock \showarticletitle{KL-UCB-switch: optimal regret bounds for stochastic
  bandits from both a distribution-dependent and a distribution-free
  viewpoints}.
\newblock \bibinfo{journal}{\emph{arXiv preprint arXiv:1805.05071}}
  (\bibinfo{year}{2018}).
\newblock


\bibitem[\protect\citeauthoryear{Garivier, Lattimore, and Kaufmann}{Garivier
  et~al\mbox{.}}{2016}]%
        {garivier2016explore}
\bibfield{author}{\bibinfo{person}{Aur{\'e}lien Garivier}, \bibinfo{person}{Tor
  Lattimore}, {and} \bibinfo{person}{Emilie Kaufmann}.}
  \bibinfo{year}{2016}\natexlab{}.
\newblock \showarticletitle{On explore-then-commit strategies}. In
  \bibinfo{booktitle}{\emph{Advances in Neural Information Processing
  Systems}}. \bibinfo{pages}{784--792}.
\newblock


\bibitem[\protect\citeauthoryear{Gomez-Uribe and Hunt}{Gomez-Uribe and
  Hunt}{2016}]%
        {gomez2016netflix}
\bibfield{author}{\bibinfo{person}{Carlos~A Gomez-Uribe} {and}
  \bibinfo{person}{Neil Hunt}.} \bibinfo{year}{2016}\natexlab{}.
\newblock \showarticletitle{The netflix recommender system: Algorithms,
  business value, and innovation}.
\newblock \bibinfo{journal}{\emph{ACM Transactions on Management Information
  Systems (TMIS)}} \bibinfo{volume}{6}, \bibinfo{number}{4}
  (\bibinfo{year}{2016}), \bibinfo{pages}{13}.
\newblock


\bibitem[\protect\citeauthoryear{Grbovic and Cheng}{Grbovic and Cheng}{2018}]%
        {grbovic2018real}
\bibfield{author}{\bibinfo{person}{Mihajlo Grbovic} {and}
  \bibinfo{person}{Haibin Cheng}.} \bibinfo{year}{2018}\natexlab{}.
\newblock \showarticletitle{Real-time personalization using embeddings for
  search ranking at Airbnb}. In \bibinfo{booktitle}{\emph{Proceedings of the
  24th ACM SIGKDD International Conference on Knowledge Discovery \& Data
  Mining}}. ACM, \bibinfo{pages}{311--320}.
\newblock


\bibitem[\protect\citeauthoryear{Gruson, Chandar, Charbuillet, McInerney,
  Hansen, Tardieu, and Carterette}{Gruson et~al\mbox{.}}{2019}]%
        {gruson2019offline}
\bibfield{author}{\bibinfo{person}{Alois Gruson}, \bibinfo{person}{Praveen
  Chandar}, \bibinfo{person}{Christophe Charbuillet}, \bibinfo{person}{James
  McInerney}, \bibinfo{person}{Samantha Hansen}, \bibinfo{person}{Damien
  Tardieu}, {and} \bibinfo{person}{Ben Carterette}.}
  \bibinfo{year}{2019}\natexlab{}.
\newblock \showarticletitle{Offline Evaluation to Make Decisions About
  PlaylistRecommendation Algorithms}. In \bibinfo{booktitle}{\emph{Proceedings
  of the Twelfth ACM International Conference on Web Search and Data Mining}}.
  ACM, \bibinfo{pages}{420--428}.
\newblock


\bibitem[\protect\citeauthoryear{Guigour{\`e}s, Ho, Koriagin, Sheikh, Bergmann,
  and Shirvany}{Guigour{\`e}s et~al\mbox{.}}{2018}]%
        {guigoures2018hierarchical}
\bibfield{author}{\bibinfo{person}{Romain Guigour{\`e}s},
  \bibinfo{person}{Yuen~King Ho}, \bibinfo{person}{Evgenii Koriagin},
  \bibinfo{person}{Abdul-Saboor Sheikh}, \bibinfo{person}{Urs Bergmann}, {and}
  \bibinfo{person}{Reza Shirvany}.} \bibinfo{year}{2018}\natexlab{}.
\newblock \showarticletitle{A hierarchical bayesian model for size
  recommendation in fashion}. In \bibinfo{booktitle}{\emph{Proceedings of the
  12th ACM Conference on Recommender Systems}}. ACM, \bibinfo{pages}{392--396}.
\newblock


\bibitem[\protect\citeauthoryear{Gunawardana and Shani}{Gunawardana and
  Shani}{2015}]%
        {gunawardana2015evaluating}
\bibfield{author}{\bibinfo{person}{Asela Gunawardana} {and}
  \bibinfo{person}{Guy Shani}.} \bibinfo{year}{2015}\natexlab{}.
\newblock \showarticletitle{Evaluating recommender systems}.
\newblock In \bibinfo{booktitle}{\emph{Recommender systems handbook}}.
  \bibinfo{publisher}{Springer}, \bibinfo{pages}{265--308}.
\newblock


\bibitem[\protect\citeauthoryear{Hill, Nassif, Liu, Iyer, and
  Vishwanathan}{Hill et~al\mbox{.}}{2017}]%
        {hill2017efficient}
\bibfield{author}{\bibinfo{person}{Daniel~N Hill}, \bibinfo{person}{Houssam
  Nassif}, \bibinfo{person}{Yi Liu}, \bibinfo{person}{Anand Iyer}, {and}
  \bibinfo{person}{SVN Vishwanathan}.} \bibinfo{year}{2017}\natexlab{}.
\newblock \showarticletitle{An efficient bandit algorithm for realtime
  multivariate optimization}. In \bibinfo{booktitle}{\emph{Proceedings of the
  23rd ACM SIGKDD International Conference on Knowledge Discovery and Data
  Mining}}. ACM, \bibinfo{pages}{1813--1821}.
\newblock


\bibitem[\protect\citeauthoryear{Isinkaye, Folajimi, and Ojokoh}{Isinkaye
  et~al\mbox{.}}{2015}]%
        {isinkaye2015recommendation}
\bibfield{author}{\bibinfo{person}{FO Isinkaye}, \bibinfo{person}{YO Folajimi},
  {and} \bibinfo{person}{BA Ojokoh}.} \bibinfo{year}{2015}\natexlab{}.
\newblock \showarticletitle{Recommendation systems: Principles, methods and
  evaluation}.
\newblock \bibinfo{journal}{\emph{Egyptian Informatics Journal}}
  \bibinfo{volume}{16}, \bibinfo{number}{3} (\bibinfo{year}{2015}),
  \bibinfo{pages}{261--273}.
\newblock


\bibitem[\protect\citeauthoryear{Johnson}{Johnson}{2014}]%
        {johnson2014logistic}
\bibfield{author}{\bibinfo{person}{Christopher~C Johnson}.}
  \bibinfo{year}{2014}\natexlab{}.
\newblock \showarticletitle{Logistic matrix factorization for implicit feedback
  data}.
\newblock \bibinfo{journal}{\emph{Advances in Neural Information Processing
  Systems}}  \bibinfo{volume}{27} (\bibinfo{year}{2014}).
\newblock


\bibitem[\protect\citeauthoryear{Kalathil, Nayyar, and Jain}{Kalathil
  et~al\mbox{.}}{2014}]%
        {kalathil2014decentralized}
\bibfield{author}{\bibinfo{person}{Dileep Kalathil}, \bibinfo{person}{Naumaan
  Nayyar}, {and} \bibinfo{person}{Rahul Jain}.}
  \bibinfo{year}{2014}\natexlab{}.
\newblock \showarticletitle{Decentralized learning for multiplayer multiarmed
  bandits}.
\newblock \bibinfo{journal}{\emph{IEEE Transactions on Information Theory}}
  \bibinfo{volume}{60}, \bibinfo{number}{4} (\bibinfo{year}{2014}),
  \bibinfo{pages}{2331--2345}.
\newblock


\bibitem[\protect\citeauthoryear{Kim and Tewari}{Kim and Tewari}{2019}]%
        {kim2019optimality}
\bibfield{author}{\bibinfo{person}{Baekjin Kim} {and} \bibinfo{person}{Ambuj
  Tewari}.} \bibinfo{year}{2019}\natexlab{}.
\newblock \showarticletitle{On the Optimality of Perturbations in Stochastic
  and Adversarial Multi-armed Bandit Problems}.
\newblock \bibinfo{journal}{\emph{arXiv preprint arXiv:1902.00610}}
  (\bibinfo{year}{2019}).
\newblock


\bibitem[\protect\citeauthoryear{Komiyama}{Komiyama}{2019}]%
        {ProductizationContextualBandit20191}
\bibfield{author}{\bibinfo{person}{Honda J. Nakagawa~H. Komiyama, J.}}
  \bibinfo{year}{2019}\natexlab{}.
\newblock \showarticletitle{{Optimal Regret Analysis of Thompson Sampling in
  Stochastic Multi-armed Bandit Problem with Multiple Plays}}.
\newblock  (\bibinfo{year}{2019}).
\newblock


\bibitem[\protect\citeauthoryear{Kreps, Narkhede, Rao, et~al\mbox{.}}{Kreps
  et~al\mbox{.}}{2011}]%
        {kreps2011kafka}
\bibfield{author}{\bibinfo{person}{Jay Kreps}, \bibinfo{person}{Neha Narkhede},
  \bibinfo{person}{Jun Rao}, {et~al\mbox{.}}} \bibinfo{year}{2011}\natexlab{}.
\newblock \showarticletitle{Kafka: A distributed messaging system for log
  processing}. In \bibinfo{booktitle}{\emph{Proceedings of the NetDB}}.
  \bibinfo{pages}{1--7}.
\newblock


\bibitem[\protect\citeauthoryear{Kveton, Szepesvari, Ghavamzadeh, and
  Boutilier}{Kveton et~al\mbox{.}}{2019}]%
        {kveton2019perturbed}
\bibfield{author}{\bibinfo{person}{Branislav Kveton}, \bibinfo{person}{Csaba
  Szepesvari}, \bibinfo{person}{Mohammad Ghavamzadeh}, {and}
  \bibinfo{person}{Craig Boutilier}.} \bibinfo{year}{2019}\natexlab{}.
\newblock \showarticletitle{Perturbed-History Exploration in Stochastic
  Multi-Armed Bandits}.
\newblock \bibinfo{journal}{\emph{arXiv preprint arXiv:1902.10089}}
  (\bibinfo{year}{2019}).
\newblock


\bibitem[\protect\citeauthoryear{Kwon, Perchet, and Vernade}{Kwon
  et~al\mbox{.}}{2017}]%
        {kwon2017sparse}
\bibfield{author}{\bibinfo{person}{Joon Kwon}, \bibinfo{person}{Vianney
  Perchet}, {and} \bibinfo{person}{Claire Vernade}.}
  \bibinfo{year}{2017}\natexlab{}.
\newblock \showarticletitle{Sparse stochastic bandits}.
\newblock \bibinfo{journal}{\emph{arXiv preprint arXiv:1706.01383}}
  (\bibinfo{year}{2017}).
\newblock


\bibitem[\protect\citeauthoryear{Lattimore}{Lattimore}{2015}]%
        {lattimore2015optimally}
\bibfield{author}{\bibinfo{person}{Tor Lattimore}.}
  \bibinfo{year}{2015}\natexlab{}.
\newblock \showarticletitle{Optimally confident UCB: Improved regret for
  finite-armed bandits}.
\newblock \bibinfo{journal}{\emph{arXiv preprint arXiv:1507.07880}}
  (\bibinfo{year}{2015}).
\newblock


\bibitem[\protect\citeauthoryear{Lattimore}{Lattimore}{2016}]%
        {lattimore2016regret}
\bibfield{author}{\bibinfo{person}{Tor Lattimore}.}
  \bibinfo{year}{2016}\natexlab{}.
\newblock \showarticletitle{Regret analysis of the anytime optimally confident
  UCB algorithm}.
\newblock \bibinfo{journal}{\emph{arXiv preprint arXiv:1603.08661}}
  (\bibinfo{year}{2016}).
\newblock


\bibitem[\protect\citeauthoryear{Le and Smola}{Le and Smola}{2007}]%
        {le2007direct}
\bibfield{author}{\bibinfo{person}{Quoc Le} {and} \bibinfo{person}{Alexander
  Smola}.} \bibinfo{year}{2007}\natexlab{}.
\newblock \showarticletitle{Direct optimization of ranking measures}.
\newblock \bibinfo{journal}{\emph{arXiv preprint arXiv:0704.3359}}
  (\bibinfo{year}{2007}).
\newblock


\bibitem[\protect\citeauthoryear{Li, Chu, Langford, Moon, and Wang}{Li
  et~al\mbox{.}}{2012}]%
        {li2012unbiased}
\bibfield{author}{\bibinfo{person}{Lihong Li}, \bibinfo{person}{Wei Chu},
  \bibinfo{person}{John Langford}, \bibinfo{person}{Taesup Moon}, {and}
  \bibinfo{person}{Xuanhui Wang}.} \bibinfo{year}{2012}\natexlab{}.
\newblock \showarticletitle{An unbiased offline evaluation of contextual bandit
  algorithms with generalized linear models}. In
  \bibinfo{booktitle}{\emph{Proceedings of the Workshop on On-line Trading of
  Exploration and Exploitation 2}}. \bibinfo{pages}{19--36}.
\newblock


\bibitem[\protect\citeauthoryear{Li, Chu, Langford, and Schapire}{Li
  et~al\mbox{.}}{2010}]%
        {li2010contextual}
\bibfield{author}{\bibinfo{person}{Lihong Li}, \bibinfo{person}{Wei Chu},
  \bibinfo{person}{John Langford}, {and} \bibinfo{person}{Robert~E Schapire}.}
  \bibinfo{year}{2010}\natexlab{}.
\newblock \showarticletitle{A contextual-bandit approach to personalized news
  article recommendation}. In \bibinfo{booktitle}{\emph{Proceedings of the 19th
  international conference on World wide web}}. ACM, \bibinfo{pages}{661--670}.
\newblock


\bibitem[\protect\citeauthoryear{Liang, Altosaar, Charlin, and Blei}{Liang
  et~al\mbox{.}}{2016}]%
        {liang2016factorization}
\bibfield{author}{\bibinfo{person}{Dawen Liang}, \bibinfo{person}{Jaan
  Altosaar}, \bibinfo{person}{Laurent Charlin}, {and} \bibinfo{person}{David~M
  Blei}.} \bibinfo{year}{2016}\natexlab{}.
\newblock \showarticletitle{Factorization meets the item embedding:
  Regularizing matrix factorization with item co-occurrence}. In
  \bibinfo{booktitle}{\emph{Proceedings of the 10th ACM conference on
  recommender systems}}. ACM, \bibinfo{pages}{59--66}.
\newblock


\bibitem[\protect\citeauthoryear{Liu, Lee, and Shroff}{Liu
  et~al\mbox{.}}{2018a}]%
        {liu2018change}
\bibfield{author}{\bibinfo{person}{Fang Liu}, \bibinfo{person}{Joohyun Lee},
  {and} \bibinfo{person}{Ness Shroff}.} \bibinfo{year}{2018}\natexlab{a}.
\newblock \showarticletitle{A change-detection based framework for
  piecewise-stationary multi-armed bandit problem}. In
  \bibinfo{booktitle}{\emph{Thirty-Second AAAI Conference on Artificial
  Intelligence}}.
\newblock


\bibitem[\protect\citeauthoryear{Liu, Wang, Buccapatnam, and Shroff}{Liu
  et~al\mbox{.}}{2018b}]%
        {liu2018ucboost}
\bibfield{author}{\bibinfo{person}{Fang Liu}, \bibinfo{person}{Sinong Wang},
  \bibinfo{person}{Swapna Buccapatnam}, {and} \bibinfo{person}{Ness Shroff}.}
  \bibinfo{year}{2018}\natexlab{b}.
\newblock \showarticletitle{UCBoost: a boosting approach to tame complexity and
  optimality for stochastic bandits}.
\newblock \bibinfo{journal}{\emph{arXiv preprint arXiv:1804.05929}}
  (\bibinfo{year}{2018}).
\newblock


\bibitem[\protect\citeauthoryear{Liu and Zhao}{Liu and Zhao}{2010}]%
        {liu2010distributed}
\bibfield{author}{\bibinfo{person}{Keqin Liu} {and} \bibinfo{person}{Qing
  Zhao}.} \bibinfo{year}{2010}\natexlab{}.
\newblock \showarticletitle{Distributed learning in multi-armed bandit with
  multiple players}.
\newblock \bibinfo{journal}{\emph{IEEE Transactions on Signal Processing}}
  \bibinfo{volume}{58}, \bibinfo{number}{11} (\bibinfo{year}{2010}),
  \bibinfo{pages}{5667--5681}.
\newblock


\bibitem[\protect\citeauthoryear{Lomas, Forlizzi, Poonwala, Patel, Shodhan,
  Patel, Koedinger, and Brunskill}{Lomas et~al\mbox{.}}{2016}]%
        {lomas2016interface}
\bibfield{author}{\bibinfo{person}{J~Derek Lomas}, \bibinfo{person}{Jodi
  Forlizzi}, \bibinfo{person}{Nikhil Poonwala}, \bibinfo{person}{Nirmal Patel},
  \bibinfo{person}{Sharan Shodhan}, \bibinfo{person}{Kishan Patel},
  \bibinfo{person}{Ken Koedinger}, {and} \bibinfo{person}{Emma Brunskill}.}
  \bibinfo{year}{2016}\natexlab{}.
\newblock \showarticletitle{Interface design optimization as a multi-armed
  bandit problem}. In \bibinfo{booktitle}{\emph{Proceedings of the 2016 CHI
  Conference on Human Factors in Computing Systems}}. ACM,
  \bibinfo{pages}{4142--4153}.
\newblock


\bibitem[\protect\citeauthoryear{Lugosi and Mehrabian}{Lugosi and
  Mehrabian}{2018}]%
        {lugosi2018multiplayer}
\bibfield{author}{\bibinfo{person}{G{\'a}bor Lugosi} {and}
  \bibinfo{person}{Abbas Mehrabian}.} \bibinfo{year}{2018}\natexlab{}.
\newblock \showarticletitle{Multiplayer bandits without observing collision
  information}.
\newblock \bibinfo{journal}{\emph{arXiv preprint arXiv:1808.08416}}
  (\bibinfo{year}{2018}).
\newblock


\bibitem[\protect\citeauthoryear{Mao, Chen, Wagle, Pan, Natkovich, and
  Matheson}{Mao et~al\mbox{.}}{2018}]%
        {mao2018batched}
\bibfield{author}{\bibinfo{person}{Yizhi Mao}, \bibinfo{person}{Miao Chen},
  \bibinfo{person}{Abhinav Wagle}, \bibinfo{person}{Junwei Pan},
  \bibinfo{person}{Michael Natkovich}, {and} \bibinfo{person}{Don Matheson}.}
  \bibinfo{year}{2018}\natexlab{}.
\newblock \showarticletitle{A Batched Multi-Armed Bandit Approach to News
  Headline Testing}. In \bibinfo{booktitle}{\emph{2018 IEEE International
  Conference on Big Data (Big Data)}}. IEEE, \bibinfo{pages}{1966--1973}.
\newblock


\bibitem[\protect\citeauthoryear{McConachie and Berenson}{McConachie and
  Berenson}{2017}]%
        {mcconachie2017bandit}
\bibfield{author}{\bibinfo{person}{Dale McConachie} {and}
  \bibinfo{person}{Dmitry Berenson}.} \bibinfo{year}{2017}\natexlab{}.
\newblock \showarticletitle{Bandit-based model selection for deformable object
  manipulation}.
\newblock \bibinfo{journal}{\emph{arXiv preprint arXiv:1703.10254}}
  (\bibinfo{year}{2017}).
\newblock


\bibitem[\protect\citeauthoryear{McInerney, Lacker, Hansen, Higley, Bouchard,
  Gruson, and Mehrotra}{McInerney et~al\mbox{.}}{2018}]%
        {mcinerney2018explore}
\bibfield{author}{\bibinfo{person}{James McInerney}, \bibinfo{person}{Benjamin
  Lacker}, \bibinfo{person}{Samantha Hansen}, \bibinfo{person}{Karl Higley},
  \bibinfo{person}{Hugues Bouchard}, \bibinfo{person}{Alois Gruson}, {and}
  \bibinfo{person}{Rishabh Mehrotra}.} \bibinfo{year}{2018}\natexlab{}.
\newblock \showarticletitle{Explore, exploit, and explain: personalizing
  explainable recommendations with bandits}. In
  \bibinfo{booktitle}{\emph{Proceedings of the 12th ACM Conference on
  Recommender Systems}}. ACM, \bibinfo{pages}{31--39}.
\newblock


\bibitem[\protect\citeauthoryear{Mitsoulis-Ntompos, Hejazinia, Zhang, and
  Brady}{Mitsoulis-Ntompos et~al\mbox{.}}{2019}]%
        {mitsoulis2019simple}
\bibfield{author}{\bibinfo{person}{Pavlos Mitsoulis-Ntompos},
  \bibinfo{person}{Meisam Hejazinia}, \bibinfo{person}{Serena Zhang}, {and}
  \bibinfo{person}{Travis Brady}.} \bibinfo{year}{2019}\natexlab{}.
\newblock \showarticletitle{A Simple Deep Personalized Recommendation System}.
\newblock \bibinfo{journal}{\emph{arXiv preprint arXiv:1906.11336}}
  (\bibinfo{year}{2019}).
\newblock


\bibitem[\protect\citeauthoryear{Mnih and Salakhutdinov}{Mnih and
  Salakhutdinov}{2008}]%
        {mnih2008probabilistic}
\bibfield{author}{\bibinfo{person}{Andriy Mnih} {and} \bibinfo{person}{Ruslan~R
  Salakhutdinov}.} \bibinfo{year}{2008}\natexlab{}.
\newblock \showarticletitle{Probabilistic matrix factorization}. In
  \bibinfo{booktitle}{\emph{Advances in neural information processing
  systems}}. \bibinfo{pages}{1257--1264}.
\newblock


\bibitem[\protect\citeauthoryear{Moulines}{Moulines}{1985}]%
        {moulines19858}
\bibfield{author}{\bibinfo{person}{Eric Moulines}.}
  \bibinfo{year}{1985}\natexlab{}.
\newblock \showarticletitle{8 On Upper-Confidence Bound Policies for
  Non-Stationary Bandit Problems}.
\newblock  (\bibinfo{year}{1985}).
\newblock


\bibitem[\protect\citeauthoryear{Oren~Barkan}{Oren~Barkan}{2017}]%
        {Oren2017Item2Vec}
\bibfield{author}{\bibinfo{person}{Noam~Koenigstein Oren~Barkan}.}
  \bibinfo{year}{2017}\natexlab{}.
\newblock \showarticletitle{{Item2Vec: Neural Item Embedding for Collaborative
  Filtering}}.
\newblock  (\bibinfo{year}{2017}).
\newblock


\bibitem[\protect\citeauthoryear{Pennington, Socher, and Manning}{Pennington
  et~al\mbox{.}}{2014}]%
        {pennington2014glove}
\bibfield{author}{\bibinfo{person}{Jeffrey Pennington},
  \bibinfo{person}{Richard Socher}, {and} \bibinfo{person}{Christopher
  Manning}.} \bibinfo{year}{2014}\natexlab{}.
\newblock \showarticletitle{Glove: Global vectors for word representation}. In
  \bibinfo{booktitle}{\emph{Proceedings of the 2014 conference on empirical
  methods in natural language processing (EMNLP)}}.
  \bibinfo{pages}{1532--1543}.
\newblock


\bibitem[\protect\citeauthoryear{Peska and Vojtas}{Peska and Vojtas}{2018}]%
        {peska2018off}
\bibfield{author}{\bibinfo{person}{Ladislav Peska} {and} \bibinfo{person}{Peter
  Vojtas}.} \bibinfo{year}{2018}\natexlab{}.
\newblock \showarticletitle{Off-line vs. On-line Evaluation of Recommender
  Systems in Small E-commerce}.
\newblock \bibinfo{journal}{\emph{arXiv preprint arXiv:1809.03186}}
  (\bibinfo{year}{2018}).
\newblock


\bibitem[\protect\citeauthoryear{Raj and Kalyani}{Raj and Kalyani}{2017}]%
        {raj2017taming}
\bibfield{author}{\bibinfo{person}{Vishnu Raj} {and} \bibinfo{person}{Sheetal
  Kalyani}.} \bibinfo{year}{2017}\natexlab{}.
\newblock \showarticletitle{Taming non-stationary bandits: A Bayesian
  approach}.
\newblock \bibinfo{journal}{\emph{arXiv preprint arXiv:1707.09727}}
  (\bibinfo{year}{2017}).
\newblock


\bibitem[\protect\citeauthoryear{Rendle, Freudenthaler, and
  Schmidt-Thieme}{Rendle et~al\mbox{.}}{2010}]%
        {rendle2010factorizing}
\bibfield{author}{\bibinfo{person}{Steffen Rendle}, \bibinfo{person}{Christoph
  Freudenthaler}, {and} \bibinfo{person}{Lars Schmidt-Thieme}.}
  \bibinfo{year}{2010}\natexlab{}.
\newblock \showarticletitle{Factorizing personalized markov chains for
  next-basket recommendation}. In \bibinfo{booktitle}{\emph{Proceedings of the
  19th international conference on World wide web}}. ACM,
  \bibinfo{pages}{811--820}.
\newblock


\bibitem[\protect\citeauthoryear{Rendle, Zhang, and Koren}{Rendle
  et~al\mbox{.}}{2019}]%
        {rendle2019difficulty}
\bibfield{author}{\bibinfo{person}{Steffen Rendle}, \bibinfo{person}{Li Zhang},
  {and} \bibinfo{person}{Yehuda Koren}.} \bibinfo{year}{2019}\natexlab{}.
\newblock \showarticletitle{On the Difficulty of Evaluating Baselines: A Study
  on Recommender Systems}.
\newblock \bibinfo{journal}{\emph{arXiv preprint arXiv:1905.01395}}
  (\bibinfo{year}{2019}).
\newblock


\bibitem[\protect\citeauthoryear{Rosenski, Shamir, and Szlak}{Rosenski
  et~al\mbox{.}}{2016}]%
        {rosenski2016multi}
\bibfield{author}{\bibinfo{person}{Jonathan Rosenski}, \bibinfo{person}{Ohad
  Shamir}, {and} \bibinfo{person}{Liran Szlak}.}
  \bibinfo{year}{2016}\natexlab{}.
\newblock \showarticletitle{Multi-player bandits--a musical chairs approach}.
  In \bibinfo{booktitle}{\emph{International Conference on Machine Learning}}.
  \bibinfo{pages}{155--163}.
\newblock


\bibitem[\protect\citeauthoryear{Ross, Mineiro, and Langford}{Ross
  et~al\mbox{.}}{2013}]%
        {ross2013normalized}
\bibfield{author}{\bibinfo{person}{St{\'e}phane Ross}, \bibinfo{person}{Paul
  Mineiro}, {and} \bibinfo{person}{John Langford}.}
  \bibinfo{year}{2013}\natexlab{}.
\newblock \showarticletitle{Normalized online learning}.
\newblock \bibinfo{journal}{\emph{arXiv preprint arXiv:1305.6646}}
  (\bibinfo{year}{2013}).
\newblock


\bibitem[\protect\citeauthoryear{Scott}{Scott}{2015}]%
        {scott2015multi}
\bibfield{author}{\bibinfo{person}{Steven~L Scott}.}
  \bibinfo{year}{2015}\natexlab{}.
\newblock \showarticletitle{Multi-armed bandit experiments in the online
  service economy}.
\newblock \bibinfo{journal}{\emph{Applied Stochastic Models in Business and
  Industry}} \bibinfo{volume}{31}, \bibinfo{number}{1} (\bibinfo{year}{2015}),
  \bibinfo{pages}{37--45}.
\newblock


\bibitem[\protect\citeauthoryear{Seldin and Lugosi}{Seldin and Lugosi}{2017}]%
        {seldin2017improved}
\bibfield{author}{\bibinfo{person}{Yevgeny Seldin} {and}
  \bibinfo{person}{G{\'a}bor Lugosi}.} \bibinfo{year}{2017}\natexlab{}.
\newblock \showarticletitle{An improved parametrization and analysis of the
  EXP3++ algorithm for stochastic and adversarial bandits}.
\newblock \bibinfo{journal}{\emph{arXiv preprint arXiv:1702.06103}}
  (\bibinfo{year}{2017}).
\newblock


\bibitem[\protect\citeauthoryear{Shipra~Agrawal}{Shipra~Agrawal}{2012}]%
        {shipra2012thompson}
\bibfield{author}{\bibinfo{person}{Navin~Goyal Shipra~Agrawal}.}
  \bibinfo{year}{2012}\natexlab{}.
\newblock \showarticletitle{Analysis of Thompson Sampling for the Multi-armed
  Bandit Problem}. In \bibinfo{booktitle}{\emph{25th Annual Conference on
  Learning Theory}}. \bibinfo{pages}{39.1–39.26}.
\newblock


\bibitem[\protect\citeauthoryear{Simon}{Simon}{1972}]%
        {simon1972theories}
\bibfield{author}{\bibinfo{person}{Herbert~A Simon}.}
  \bibinfo{year}{1972}\natexlab{}.
\newblock \showarticletitle{Theories of bounded rationality}.
\newblock \bibinfo{journal}{\emph{Decision and organization}}
  \bibinfo{volume}{1}, \bibinfo{number}{1} (\bibinfo{year}{1972}),
  \bibinfo{pages}{161--176}.
\newblock


\bibitem[\protect\citeauthoryear{Singla, Hassani, and Krause}{Singla
  et~al\mbox{.}}{2017}]%
        {singla2017learning}
\bibfield{author}{\bibinfo{person}{Adish Singla}, \bibinfo{person}{Hamed
  Hassani}, {and} \bibinfo{person}{Andreas Krause}.}
  \bibinfo{year}{2017}\natexlab{}.
\newblock \showarticletitle{Learning to Use Learners' Advice}.
\newblock \bibinfo{journal}{\emph{arXiv preprint arXiv:1702.04825}}
  (\bibinfo{year}{2017}).
\newblock


\bibitem[\protect\citeauthoryear{Sirotkin}{Sirotkin}{2013}]%
        {sirotkin2013search}
\bibfield{author}{\bibinfo{person}{Pavel Sirotkin}.}
  \bibinfo{year}{2013}\natexlab{}.
\newblock \showarticletitle{On search engine evaluation metrics}.
\newblock \bibinfo{journal}{\emph{arXiv preprint arXiv:1302.2318}}
  (\bibinfo{year}{2013}).
\newblock


\bibitem[\protect\citeauthoryear{Smith and Linden}{Smith and Linden}{2017}]%
        {smith2017two}
\bibfield{author}{\bibinfo{person}{Brent Smith} {and} \bibinfo{person}{Greg
  Linden}.} \bibinfo{year}{2017}\natexlab{}.
\newblock \showarticletitle{Two decades of recommender systems at Amazon. com}.
\newblock \bibinfo{journal}{\emph{Ieee internet computing}}
  \bibinfo{volume}{21}, \bibinfo{number}{3} (\bibinfo{year}{2017}),
  \bibinfo{pages}{12--18}.
\newblock


\bibitem[\protect\citeauthoryear{Tang, Jiang, Li, and Li}{Tang
  et~al\mbox{.}}{2014}]%
        {tang2014ensemble}
\bibfield{author}{\bibinfo{person}{Liang Tang}, \bibinfo{person}{Yexi Jiang},
  \bibinfo{person}{Lei Li}, {and} \bibinfo{person}{Tao Li}.}
  \bibinfo{year}{2014}\natexlab{}.
\newblock \showarticletitle{Ensemble contextual bandits for personalized
  recommendation}. In \bibinfo{booktitle}{\emph{Proceedings of the 8th ACM
  Conference on Recommender Systems}}. ACM, \bibinfo{pages}{73--80}.
\newblock


\bibitem[\protect\citeauthoryear{Vasile, Smirnova, and Conneau}{Vasile
  et~al\mbox{.}}{2016}]%
        {vasile2016meta}
\bibfield{author}{\bibinfo{person}{Flavian Vasile}, \bibinfo{person}{Elena
  Smirnova}, {and} \bibinfo{person}{Alexis Conneau}.}
  \bibinfo{year}{2016}\natexlab{}.
\newblock \showarticletitle{Meta-prod2vec: Product embeddings using
  side-information for recommendation}. In
  \bibinfo{booktitle}{\emph{Proceedings of the 10th ACM Conference on
  Recommender Systems}}. ACM, \bibinfo{pages}{225--232}.
\newblock


\bibitem[\protect\citeauthoryear{Wan, Lan, Wang, Guo, Xu, and Cheng}{Wan
  et~al\mbox{.}}{2015}]%
        {wan2015next}
\bibfield{author}{\bibinfo{person}{Shengxian Wan}, \bibinfo{person}{Yanyan
  Lan}, \bibinfo{person}{Pengfei Wang}, \bibinfo{person}{Jiafeng Guo},
  \bibinfo{person}{Jun Xu}, {and} \bibinfo{person}{Xueqi Cheng}.}
  \bibinfo{year}{2015}\natexlab{}.
\newblock \showarticletitle{Next Basket Recommendation with Neural Networks.}.
  In \bibinfo{booktitle}{\emph{RecSys Posters}}.
\newblock


\bibitem[\protect\citeauthoryear{Wang, Huang, Zhao, Zhang, Zhao, and Lee}{Wang
  et~al\mbox{.}}{2018}]%
        {wang2018billion}
\bibfield{author}{\bibinfo{person}{Jizhe Wang}, \bibinfo{person}{Pipei Huang},
  \bibinfo{person}{Huan Zhao}, \bibinfo{person}{Zhibo Zhang},
  \bibinfo{person}{Binqiang Zhao}, {and} \bibinfo{person}{Dik~Lun Lee}.}
  \bibinfo{year}{2018}\natexlab{}.
\newblock \showarticletitle{Billion-scale commodity embedding for e-commerce
  recommendation in alibaba}. In \bibinfo{booktitle}{\emph{Proceedings of the
  24th ACM SIGKDD International Conference on Knowledge Discovery \& Data
  Mining}}. ACM, \bibinfo{pages}{839--848}.
\newblock


\bibitem[\protect\citeauthoryear{Wang, Cao, and Wang}{Wang
  et~al\mbox{.}}{2019}]%
        {wang2019survey}
\bibfield{author}{\bibinfo{person}{Shoujin Wang}, \bibinfo{person}{Longbing
  Cao}, {and} \bibinfo{person}{Yan Wang}.} \bibinfo{year}{2019}\natexlab{}.
\newblock \showarticletitle{A Survey on Session-based Recommender Systems}.
\newblock \bibinfo{journal}{\emph{arXiv preprint arXiv:1902.04864}}
  (\bibinfo{year}{2019}).
\newblock


\bibitem[\protect\citeauthoryear{Wang, Wang, and Powell}{Wang
  et~al\mbox{.}}{2016}]%
        {wang2016knowledge}
\bibfield{author}{\bibinfo{person}{Yingfei Wang}, \bibinfo{person}{Chu Wang},
  {and} \bibinfo{person}{Warren Powell}.} \bibinfo{year}{2016}\natexlab{}.
\newblock \showarticletitle{The knowledge gradient for sequential decision
  making with stochastic binary feedbacks}. In
  \bibinfo{booktitle}{\emph{International Conference on Machine Learning}}.
  \bibinfo{pages}{1138--1147}.
\newblock


\bibitem[\protect\citeauthoryear{Wei and Srivatsva}{Wei and Srivatsva}{2018}]%
        {wei2018abruptly}
\bibfield{author}{\bibinfo{person}{Lai Wei} {and} \bibinfo{person}{Vaibhav
  Srivatsva}.} \bibinfo{year}{2018}\natexlab{}.
\newblock \showarticletitle{On abruptly-changing and slowly-varying multiarmed
  bandit problems}. In \bibinfo{booktitle}{\emph{2018 Annual American Control
  Conference (ACC)}}. IEEE, \bibinfo{pages}{6291--6296}.
\newblock


\bibitem[\protect\citeauthoryear{Wu, Tang, Zhu, Wang, Xie, and Tan}{Wu
  et~al\mbox{.}}{2018}]%
        {wu2018session}
\bibfield{author}{\bibinfo{person}{Shu Wu}, \bibinfo{person}{Yuyuan Tang},
  \bibinfo{person}{Yanqiao Zhu}, \bibinfo{person}{Liang Wang},
  \bibinfo{person}{Xing Xie}, {and} \bibinfo{person}{Tieniu Tan}.}
  \bibinfo{year}{2018}\natexlab{}.
\newblock \showarticletitle{Session-based Recommendation with Graph Neural
  Networks}.
\newblock \bibinfo{journal}{\emph{arXiv preprint arXiv:1811.00855}}
  (\bibinfo{year}{2018}).
\newblock


\bibitem[\protect\citeauthoryear{Zeng, Wang, Mokhtari, and Li}{Zeng
  et~al\mbox{.}}{2016}]%
        {zeng2016online}
\bibfield{author}{\bibinfo{person}{Chunqiu Zeng}, \bibinfo{person}{Qing Wang},
  \bibinfo{person}{Shekoofeh Mokhtari}, {and} \bibinfo{person}{Tao Li}.}
  \bibinfo{year}{2016}\natexlab{}.
\newblock \showarticletitle{Online context-aware recommendation with time
  varying multi-armed bandit}. In \bibinfo{booktitle}{\emph{Proceedings of the
  22nd ACM SIGKDD international conference on Knowledge discovery and data
  mining}}. ACM, \bibinfo{pages}{2025--2034}.
\newblock


\bibitem[\protect\citeauthoryear{Zimmert and Seldin}{Zimmert and
  Seldin}{2018}]%
        {zimmert2018optimal}
\bibfield{author}{\bibinfo{person}{Julian Zimmert} {and}
  \bibinfo{person}{Yevgeny Seldin}.} \bibinfo{year}{2018}\natexlab{}.
\newblock \showarticletitle{An optimal algorithm for stochastic and adversarial
  bandits}.
\newblock \bibinfo{journal}{\emph{arXiv preprint arXiv:1807.07623}}
  (\bibinfo{year}{2018}).
\newblock


\end{thebibliography}










\end{document}